\documentclass[lettersize,journal]{IEEEtran}
\usepackage{amsmath,amsfonts}
\usepackage{algorithmic}
\usepackage{algorithm}
\usepackage{array}
\usepackage[caption=false,font=scriptsize,labelfont=rm,textfont=rm]{subfig}
\usepackage{textcomp}
\usepackage{stfloats}
\usepackage{url}
\usepackage{verbatim}
\usepackage{graphicx}
\usepackage{cite}
\usepackage{bm}			
\usepackage{amssymb}		
\allowdisplaybreaks

\usepackage[colorlinks=true,citecolor=blue,linkcolor=blue,urlcolor=blue]{hyperref}

\begin{document}

\title{\huge Over-the-Air Computation via 2D Movable Antenna Array}

\author{Nianzu Li,~Peiran Wu,~Boyu Ning,~Lipeng Zhu, and Weidong Mei
\thanks{Nianzu Li and Peiran Wu are with the School of Electronics and Information Technology, Sun Yat-sen University, Guangzhou 510006, China (e-mail: linz5@mail2.sysu.edu.cn; wupr3@mail.sysu.edu.cn).} 
\thanks{Boyu Ning and Weidong Mei are with National Key Laboratory of Wireless Communications, University of Electronic
Science and Technology of China, Chengdu 611731, China (e-mails: boydning@outlook.com; wmei@uestc.edu.cn).}
\thanks{Lipeng Zhu is with the Department of Electrical and Computer
Engineering, National University of Singapore, Singapore 117583 (e-mail: zhulp@nus.edu.sg).}}

\markboth{Journal of \LaTeX\ Class Files,~Vol.~18, No.~9, September~2020}%
{Shell \MakeLowercase{\textit{et al.}}: A Sample Article Using IEEEtran.cls for IEEE Journals}


\maketitle

\begin{abstract}
Movable antenna (MA) has emerged as a promising technology for improving the performance
of wireless communication systems, which enables local movement of the antennas to create more favorable channel conditions. In this letter, we advance its application for over-the-air computation (AirComp) network, where an access point is equipped with a two-dimensional (2D) MA array to aggregate wireless data from massive users. We aim to minimize the computation mean square error (CMSE) by jointly optimizing the antenna position vector (APV), the receive combining vector at the access point and the transmit coefficients from all users. To tackle this highly non-convex problem, we propose a two-loop iterative algorithm, where the particle swarm optimization (PSO) approach is leveraged to obtain a suboptimal APV in the outer loop while the receive combining vector and transmit coefficients are alternately optimized in the inner loop. Numerical results demonstrate that the proposed MA-enhanced AirComp network outperforms the conventional network with fixed-position antennas (FPAs). 

\end{abstract}


\section{Introduction}
\IEEEPARstart{U}{nder} the rapid development of Internet of Things (IoT), Big Data and Artificial Intelligence (AI), the demand for efficient data collection from a large number of smart users and devices has become increasingly pronounced. In response, the technique of over-the-air computation (AirComp)\cite{ref19},\cite{ref22} has been developed to enable ultra-fast wireless data aggregation by leveraging the waveform superposition property of wireless channels. In particular, AirComp can efficiently integrate communication and computation processes for real-time data collection, especially in the scenario with a large amount of IoT devices and limited spectrum resources. 

However, the computation mean square error (CMSE) of AirComp can be critically affected by the quality of wireless channels. To mitigate this limitation, the reconfigurable intelligent surface (RIS) and amplify and forward (AF) relay were exploited in the AirComp network to enhance the computation performance\cite{ref20},\cite{ref30},\cite{ref23}. The main difference between them is that, compared with AF relays, RIS can improve the radio propagation in a more energy-efficient manner through passive reflections without introducing the noise propagation. Besides, a new antenna technology, namely movable antenna (MA)\cite{ref1}, also known as fluid antenna\cite{ref16}, was also recently proposed for improving the quality of wireless channels, thus anticipated to enhance the AirComp performance. Specifically, MAs can enable local antenna movement within the given spatial region to obtain better
channel conditions, thereby introducing additional degrees of freedom (DoFs) to improve the wireless communication performance. Previous studies have verified the advantages of MAs in enhancing the performance of wireless communication systems compared to traditional fixed position antennas (FPAs). In \cite{ref3}, the maximum channel gain achieved by a single
receive MA was analyzed under both deterministic and stochastic channels. Results show that the MA system can achieve a higher channel gain over its FPA counterpart via the antenna movement, thus providing additional DoFs to improve the system performance. In \cite{ref4}, the MA-enhanced multiple-input multiple-output system was investigated, where the channel capacity was maximized by jointly optimizing the positions of MAs and the transmit covariance matrix. Exploiting MAs in multi-user communications was investigated in \cite{ref9}, where the total transmit power of the users was minimized by jointly optimizing the positions of their equipped MAs. Besides, MAs have also been integrated to other wireless applications, such as point-to-point transmission \cite{ref27}, flexible beamforming \cite{ref28}, and non-orthogonal multiple access (NOMA) \cite{ref2}.

Motivated by the above studies, in this letter, we investigate an MA-enhanced AirComp network, where an access point (AP) is equipped with a two-dimensional (2D) MA array to assist in wireless data aggregation. Note that the authors in \cite{ref21} have investigated the fluid antenna array enhanced AirComp, whereas only the one-dimensional (1D) array was considered under the ideal line-of-sight (LoS) channels. In comparison, we investigate in this letter a more general 2D MA array under the practical field-response channel model. Specifically, we formulate the problem of minimizing the CMSE by joint optimization of the antenna positions of MAs, the receive combining vector at the AP and the transmit coefficients at the users. Due to the non-convex nature of this problem, it is generally hard to find the optimal solution. Different from most related works, resorting to alternating optimization (AO) or successive convex approximation (SCA) to derive a locally optimal solution \cite{ref4},\cite{ref2}, we develop a two-loop iterative algorithm leveraging the particle swarm optimization (PSO). Simulation results show the
substantial improvements of our proposed scheme in reducing the CMSE compared with other benchmark schemes.


\section{System model and problem formulation}
\begin{figure}[t]
	\centering
	\includegraphics[width=0.43\textwidth]{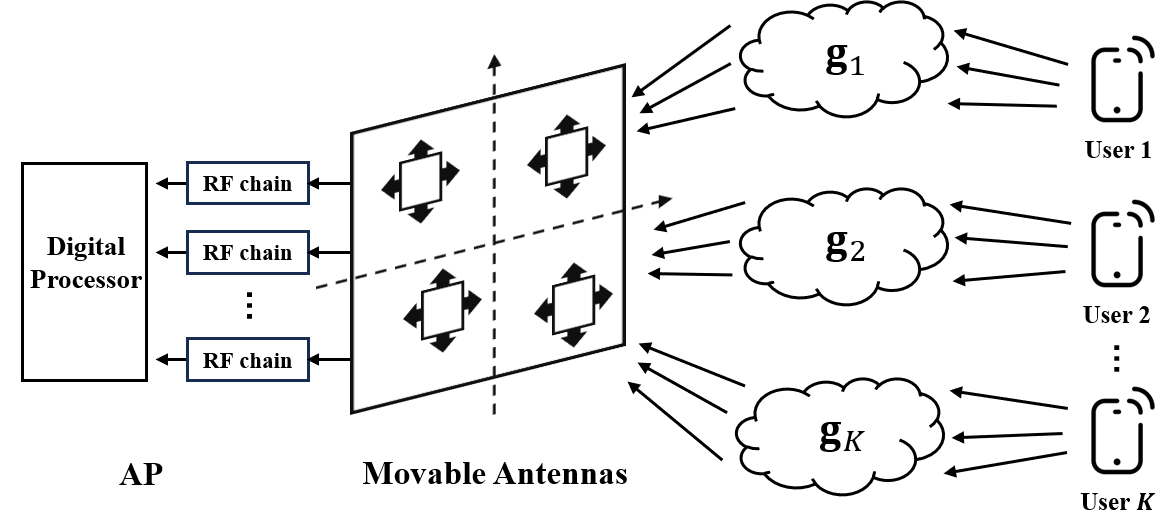}
	\caption{System model of the considered AirComp network.}
	\label{system_model}
\end{figure}

As shown in Fig. \ref{system_model}, we consider the uplink transmission of an AirComp network, which consists of $K$ single-antenna users and an AP equipped with $M$ MAs. Each MA is connected to the radio frequency (RF) chain via a flexible cable, thus can be flexibly moved in an $A\times A$ square region $\mathcal{C}_r$ at the AP for exploiting more spatial diversity gains. The position of the $m$-th MA is represented by the Cartesian coordinate, i.e., $\mathbf{r}_m=[x_m,y_m]^\mathrm{T}\in\mathcal{C}_r,1\leq m \leq M$. Let $s_k\in\mathbb{C},1\leq k \leq K$ denote the normalized information-bearing data form user $k$, with $\mathbb{E}[s_k]=0,\mathbb{E}[|s_k|^2]=1$, and $\mathbb{E}[s_{k_1}s_{k_2}^{\ast}]=0,k_1\neq k_2$. The target function that the AP aims to recover is the summation of all users' data, i.e.,
\begin{equation}
	x=\sum_{k=1}^{K}s_k.
\end{equation}
In this letter, we consider narrow-band quasi-static channels, where the users are distributed at fixed locations or with low mobility, and their surrounding propagation environment typically varies slowly. This may occur in the future IoT networks with various wireless sensors or machine-type communication (MTC) devices\cite{ref33}, such as smart cities, automated industries and smart homes. In such scenarios, MAs can be deployed at the AP to facilitate continuous data aggregation. Besides, all users are assumed to be synchronized to a common reference clock. Thus, the received signal at the AP is given by
\begin{equation}
	\mathbf{y}=\sum_{k=1}^{K}\mathbf{h}_k a_k s_k+\mathbf{n},
\end{equation}
where $\mathbf{h}_k\in\mathbb{C}^{M\times1}$ denotes the channel vector from user $k$ to the AP, $a_k$ denotes the transmit coefficient of user $k$ and $\mathbf{n}\sim\mathcal{CN}(\mathbf{0},\sigma^2\mathbf{I}_M)$ denotes the additive white Gaussian noise with zero mean and variance $\sigma^2$. The estimated function at the AP is obtained by using a digital combining vector:
\begin{equation}
	\hat{x}=\mathbf{w}^{\mathrm{H}}\mathbf{y}=\sum_{k=1}^{K}\mathbf{w}^{\mathrm{H}}\mathbf{h}_k a_k s_k+\mathbf{w}^{\mathrm{H}}\mathbf{n},
\end{equation}
where $\mathbf{w}\in\mathbb{C}^{M\times1}$ denotes the receive combining vector. 

We denote the total number of receive channel paths at the AP from user $k$, the elevation and azimuth angles of arrival (AoAs) for the $p$-th receive path from user $k$ to the AP as $L_k,1\leq k \leq K,\theta_k^p\in[0,\pi]$, and $\phi_k^p\in[0,\pi]$, respectively. Besides, it is assumed that the far-field condition is satisfied between the AP and users since the size of the moving region for MAs is much smaller than the signal propagation distance. Thus, for each user, the AoAs and amplitudes of the complex path coefficients for multiple channel paths do not change for different positions of the MAs, while only the phases of the multi-path channels vary in the MA  moving region. As such, the signal propagation distance difference of the $p$-th path for user $k$ between the position of the $m$-th MA and the origin of the receive region at the AP, $\mathbf{o}_r=[0,0]^\mathrm{T}$, is
\begin{equation}
	\rho_k^p(\mathbf{r}_m)=x_k\sin\theta_k^p\cos\phi_k^p+y_k\cos\theta_k^p.
\end{equation}
According to the field-response based channel model \cite{ref3}, the receive field-response vector (FRV) for the channel between user $k$ and the $m$-th MA at the AP is given by
\begin{equation}
	\mathbf{f}_k(\mathbf{r}_m)=\left[e^{\mathrm{j}\frac{2\pi}{\lambda}\rho_k^1(\mathbf{r}_m)},\cdots,e^{\mathrm{j}\frac{2\pi}{\lambda}\rho_k^{L_k}(\mathbf{r}_m)}\right]^\mathrm{T},
\end{equation}
where $\lambda$ denotes the carrier wavelength. Hence, the receive field-response matrix (FRM) of user $k$ is given by
\begin{equation}
	\mathbf{F}_k(\tilde{\mathbf{r}})=\left[\mathbf{f}_k(\mathbf{r}_1),\cdots,	\mathbf{f}_k(\mathbf{r}_M)\right],
	\label{eq6}
\end{equation}
with $\tilde{\mathbf{r}}=[\mathbf{r}_1^\mathrm{T},\mathbf{r}_2^\mathrm{T},\cdots,\mathbf{r}_M^\mathrm{T}]^\mathrm{T}$ denoting the antenna position vector (APV) for MAs. Finally, the channel response vector between user $k$ and the AP is expressed as
\begin{equation}
	\mathbf{h}_k(\tilde{\mathbf{r}})=\mathbf{F}_k^{\mathrm{H}}(\tilde{\mathbf{r}})\mathbf{g}_k,\label{eq7}
\end{equation}
where $\mathbf{g}_k=[g_{k,1},g_{k,2},\cdots,g_{k,L_k}]^\mathrm{T}$ is the $k$-th user’s path-response vector (PRV), representing the multi-path response coefficients from user $k$ to the original point in the receive region of the AP.

We adopt the CMSE to measure the distortion between the estimated and the target function, which is defined as
\begin{align}
	\label{eq8}
	\mathrm{CMSE}=\mathbb{E}\left[|\hat{x}-x|^2\right]=\sum_{k=1}^{K}|a_k\mathbf{w}^{\mathrm{H}}\mathbf{h}_k(\tilde{\mathbf{r}})-1|^2+\sigma^2\|\mathbf{w}\|^2.
\end{align}
The objective of this letter is to minimize the CMSE by jointly optimizing the APV for MAs, the receive combining vector, and the users' transmit coefficients. To achieve this goal, the field-response information (FRI), such as the AoAs and PRVs of the channel paths for each user, needs to be obtained. In practice, this can be acquired by using the compressed-sensing based channel estimation methods for MA systems\cite{ref6}. Let $\mathbf{a}=[a_1,a_2,\cdots,a_K]^\mathrm{T}$ denote the transmit coefficient vector, the optimization problem can be formulated as
\begin{subequations}
	\label{eq9}
	\begin{align}
		\min_{\mathbf{a},\mathbf{w},\tilde{\mathbf{r}}}& \quad \sum_{k=1}^{K}|a_k\mathbf{w}^{\mathrm{H}}\mathbf{h}_k(\tilde{\mathbf{r}})-1|^2+\sigma^2\|\mathbf{w}\|^2\label{eq9a}\\
		\mathrm{s.t.}& \quad 
		|a_k|^2\leq P_c,~1\leq k\leq K\label{eq9b},\\
		& \quad \mathbf{r}_m\in\mathcal{C}_r,~1\leq m\leq M\label{eq9c},\\
		& \quad \|\mathbf{r}_m-\mathbf{r}_n\|_2\geq D,~1\leq m\neq n\leq M\label{eq9d},
	\end{align}
\end{subequations}
where $P_c$ is the maximum power constraint for each user and $D$ is the minimum inter-MA distance $D$ to avoid the coupling effect between each antenna.

\section{Proposed solution}
Problem \eqref{eq9} is challenging to solve because the objective function is highly non-convex over $\tilde{\mathbf{r}}$ and the three high-dimensional optimization variables ($\mathbf{a},\mathbf{w},\tilde{\mathbf{r}}$) are coupled with each other. Conventional AO method cannot be directly used since it may plunge into undesired local optimum\footnote{Since there are three highly coupled optimization variables, conventional AO method may narrow the feasible space to a tiny region and converge on a locally optimal point. Moreover, due to the highly non-convexity of CMSE over $\tilde{\mathbf{r}}$, there is generally no efficient method to obtain the optimal solution. Existing optimization method, such as the SCA technique, will also result in a locally optimal solution \cite{ref4},\cite{ref2}.}. To overcome this problem, we propose a two-loop iterative algorithm to derive a suboptimal solution, where the PSO \cite{ref11} is used to optimize the APV in the outer loop while the fitness function of each particle (i.e., APV) is evaluated in the inner loop by alternately optimizing the receive combining vector and the transmit coefficient vector.

\subsection{Inner-Loop Optimization}
In the inner loop of our proposed algorithm, for each particle $\tilde{\mathbf{r}}$, we need to find the optimal receive combining vector and transmit coefficient vector to calculate its fitness value. Thus, it is formulated as the following problem
\begin{subequations}
	\label{eq10}
	\begin{align}
		\min_{\mathbf{a},\mathbf{w}}& \quad \sum_{k=1}^{K}|a_k\mathbf{w}^{\mathrm{H}}\mathbf{h}_k-1|^2+\sigma^2\|\mathbf{w}\|^2\label{eq10a}\\
		\mathrm{s.t.}& \quad 
		|a_k|^2\leq P_c,~k=1,\cdots,K\label{eq10b}.
	\end{align}
\end{subequations}
To resolve the coupling of $\mathbf{w}$ and $\mathbf{a}$, we decouple
problem \eqref{eq10} into two tractable convex sub-problems and solve them alternately until convergence.

1) Optimizing $\mathbf{w}$ with given $\mathbf{a}$: Given $\mathbf{a}$, the optimization problem is expressed as
\begin{align}
		\min_{\mathbf{w}} \quad f(\mathbf{w})=\sum_{k=1}^{K}|a_k\mathbf{w}^{\mathrm{H}}\mathbf{h}_k-1|^2+\sigma^2\|\mathbf{w}\|^2\label{eq11}.
\end{align}
Note that problem \eqref{eq11} is an unconstrained convex problem and the first-order partial derivative of the objective function with respect to $\mathbf{w}$ is
\begin{equation}
	\frac{\partial f(\mathbf{w})}{\partial\mathbf{w}}=2\left(\sum_{k=1}^{K}|a_k|^2\mathbf{h}_k\mathbf{h}_k^{\mathrm{H}}+\sigma^2\mathbf{I}\right)\mathbf{w}-2\sum_{k=1}^{K}a_k\mathbf{h}_k.
\end{equation}
Therefore, by setting $\frac{\partial f(\mathbf{w})}{\partial\mathbf{w}}=0$, we can derive the optimal $\mathbf{w}$ in closed forms, as
\begin{equation}
	\mathbf{w}^\star= \left(\sum_{k=1}^{K}|a_k|^2\mathbf{h}_k\mathbf{h}_k^{\mathrm{H}}+\sigma^2\mathbf{I}\right)^{-1}\sum_{k=1}^{K}a_k\mathbf{h}_k.\label{eq13}
\end{equation}

2) Optimizing $\mathbf{a}$ with given $\mathbf{w}$: Given $\mathbf{w}$, the optimization problem is expressed as
\begin{subequations}
	\label{eq14}
	\begin{align}
		\min_{\mathbf{a}}& \quad \sum_{k=1}^{K}|a_k\mathbf{w}^{\mathrm{H}}\mathbf{h}_k-1|^2\label{eq14a}\\
		\mathrm{s.t.}& \quad 
		|a_k|^2\leq P_c,~k=1,\cdots,K\label{eq14b}.
	\end{align}
\end{subequations}
To obtain the optimal transmit coefficient vector, we provide the following proposition.

\textit{Proposition 1}: Denote $b_k\triangleq\mathbf{w}^{\mathrm{H}}\mathbf{h}_k=|b_k|e^{\mathrm{j}\angle b_k}$, with amplitude $|b_k|$ and phase $\angle b_k$. Then, the closed-form optimal solution for problem \eqref{eq14} is
\begin{equation}
	a_k^{\star}=\min\left(\sqrt{P_c},\frac{1}{|b_k|}\right)e^{-\mathrm{j}\angle b_k}.\label{eq15}
\end{equation}

\textit{Proof}: It is noted that in the objective function of problem \eqref{eq14}, each summation term is only related with a specific $a_k$. Thus, for any $k$, if $|b_k|\geq1/\sqrt{P_c}$, we can always choose $a_k^\star=1/|b_k|e^{-\mathrm{j}\angle b_k}$ such that $|a_k\mathbf{w}^{\mathrm{H}}\mathbf{h}_k-1|$ is equal to zero. Besides, if $|b_k|<1/\sqrt{P_c}$, the optimal $a_k$ can be obtained by choosing the maximum absolute value $\sqrt{P_c}$ and the reverse phase of $b_k$, i.e., $a_k^\star=\sqrt{P_c}e^{-\mathrm{j}\angle b_k}$. Therefore, we can derive the optimal $a_k$ in closed forms, as
$a_k^{\star}=\min(\sqrt{P_c},\frac{1}{|b_k|})e^{-\mathrm{j}\angle b_k}$. This thus proves Proposition 1. $\hfill\blacksquare$

Therefore, for each particle, its corresponding receive combining vector and transmit coefficient vector can be alternately optimized until convergence. Subsequently, the CMSE can be expressed as a function of the APV $\tilde{\mathbf{r}}$, i.e.,
\begin{equation}
	\mathrm{CMSE}(\tilde{\mathbf{r}})=\sum_{k=1}^{K}|a_k^{\star}{\mathbf{w}^{\star}}^{\mathrm{H}}\mathbf{h}_k(\tilde{\mathbf{r}})-1|^2+\sigma^2\|\mathbf{w}^{\star}\|^2.\label{eq16}
\end{equation}

\subsection{Outer-Loop Optimization}
In the outer loop of our proposed algorithm, we utilize the PSO method as an efficient approach to optimize the APV $\tilde{\mathbf{r}}$. First, $N$ particles are randomly generated as a particle swarm with positions
\begin{align}
	\tilde{\mathbf{r}}_n^{(0)}=[\underbrace{x_{n,1}^{(0)},y_{n,1}^{(0)}}_{\text{MA $1$}},\underbrace{x_{n,2}^{(0)},y_{n,2}^{(0)}}_{\text{MA $2$}},\cdots,\underbrace{x_{n,M}^{(0)},y_{n,M}^{(0)}}_{\text{MA $M$}}]^\mathrm{T},
\end{align}
and velocities
\begin{align}
	\tilde{\mathbf{v}}_n^{(0)}=[\underbrace{v_{n,x_1}^{(0)},v_{n,y_1}^{(0)}}_{\text{MA $1$}},\underbrace{v_{n,x_2}^{(0)},v_{n,y_2}^{(0)}}_{\text{MA $2$}},\cdots,\underbrace{v_{n,x_M}^{(0)},v_{n,y_M}^{(0)}}_{\text{MA $M$}}]^\mathrm{T},
\end{align}
where $x_{n,m}^{(0)},y_{n,m}^{(0)}\sim \mathcal{U}[-A/2,A/2]$ for $1\leq n \leq N, 1\leq m \leq M$ ensures that the initial position of each MA satisfies constraint \eqref{eq9c}. For each particle, the fitness value can be evaluated in the inner-loop optimization, given by \eqref{eq16}. Furthermore, in order to ensure the minimum inter-MA distance, i.e., constraint \eqref{eq9d}, we introduce a penalty factor to the fitness function, i.e.,
\begin{equation}
	\scalebox{0.9}{$
	\mathcal{F}\left(\tilde{\mathbf{r}}_n^{(t)}\right)=\mathrm{CMSE}\left(\tilde{\mathbf{r}}_n^{(t)}\right)+\tau\left|\mathcal{P}\left(\tilde{\mathbf{r}}_n^{(t)}\right)\right|,$}\label{eq19}
\end{equation}
where $t$ denotes the iteration index, $\mathcal{P}\left(\tilde{\mathbf{r}}_n^{(t)}\right)\triangleq\{(\mathbf{r}_i,\mathbf{r}_j)|\| \mathbf{r}_i-\mathbf{r}_j\|_2< D,1\leq i< j \leq M \}$ and $\tau$ is a large positive penalty parameter such that $\tau>\mathrm{CMSE}\left(\tilde{\mathbf{r}}_n^{(t)}\right)$ holds for all APVs. Thus, during the iterations, the particles can be driven to move to positions where the minimum inter-MA distance is satisfied, i.e., $\left|\mathcal{P}\left(\tilde{\mathbf{r}}_n^{(t)}\right)\right|=0$. 

In each iteration, according to \eqref{eq19}, we can update the personal best position $\tilde{\mathbf{r}}_{n,pbest}$ of each particle and the global best position $\tilde{\mathbf{r}}_{gbest}$ of all particles. Then, based on the principle of PSO, the velocity and position of each particle are updated as
\begin{align}
	\scalebox{0.9}{$
	\tilde{\mathbf{v}}_n^{(t+1)}=\omega\tilde{\mathbf{v}}_n^{(t)}+$}&\scalebox{0.96}{$c_1\alpha_1\left(\tilde{\mathbf{r}}_{n,pbest}-\tilde{\mathbf{r}}_n^{(t)}\right)+c_2\alpha_2\left(\tilde{\mathbf{r}}_{gbest}-\tilde{\mathbf{r}}_n^{(t)}\right)$},\label{eq20}\\
	&\scalebox{0.9}{$\tilde{\mathbf{r}}_n^{(t+1)}=\mathcal{G}\left(\tilde{\mathbf{r}}_n^{(t)}+\tilde{\mathbf{v}}_n^{(t+1)}\right)$},\label{eq21}
\end{align}
where $c_1$ and $c_2$ are the personal and global learning factors, $\alpha_1$ and $\alpha_2$ are two random parameters uniformly distributed over $[0,1]$, and $\omega$ is the inertia weight of the particle search. In particular, to balance the accuracy and convergence speed (exploration and exploitation) of the PSO, a linearly decreasing inertia weight is adopted\cite{ref34}, i.e., $\omega=\omega_{\max}-(\omega_{\max}-\omega_{\min})t/T$, where $\omega_{\max}$ and $\omega_{\min}$ are the predefined upper and lower bounds of $\omega$, and $T$ is the maximum iteration number. Meanwhile, to guarantee that each MA does not exceed the finite region $\mathcal{C}_r$, $\tilde{\mathbf{r}}_n^{(t+1)}$ is confined within the given minimum and maximum value, i.e.,
\begin{equation}	
	\scalebox{0.9}{$[\mathcal{G}\left(\tilde{\mathbf{r}}\right)]_i=
	\begin{cases}
		-\dfrac{A}{2};&{\text{if}}\ [\tilde{\mathbf{r}}]_i<-\dfrac{A}{2}, \\
		\dfrac{A}{2};&{\text{if}}\ [\tilde{\mathbf{r}}]_i>\dfrac{A}{2}, \\
		[\tilde{\mathbf{r}}]_i;&{\text{otherwise}}.	
	\end{cases}$}
\end{equation}
Finally, with each particle updating its personal optimal position iteratively based on its own experience and the sharing experience from the swarm, the global best position among all the particles can be updated with its fitness value non-increasing until convergence. Thus, we can obtain a suboptimal solution for the APV.

\subsection{Overall Algorithm}
The overall algorithm for solving problem \eqref{eq9} is presented in \textbf{Algorithm \ref{alg1}}. Note that the convergence of this algorithm is guaranteed since the objective function is non-increasing over the iterations and has a lower bound of zero. Besides, in the inner-loop, the
computational complexities for calculating $\mathbf{w}^\star$ and $\mathbf{a}^\star$ are about $\mathcal{O}(M^3)$ and $\mathcal{O}(KM)$, respectively. Thus, the complexity of the inner-loop optimization is $\mathcal{O}(J(M^3+KM))$, where $J$ is the corresponding number of iterations. In the outer-loop, the complexity for calculating $\tilde{\mathbf{r}}_n^{(t)}$ and $\tilde{\mathbf{v}}_n^{(t)}$  is $\mathcal{O}(M)$. Therefore, the overall complexity of our proposed algorithm is $\mathcal{O}\left(NT(JM^3+JKM+M)\right)$.
\begin{algorithm}[t]
	\footnotesize
	\caption{Proposed two-loop iterative Algorithm for solving Problem \eqref{eq9}.}
	\begin{algorithmic}[1]
		\STATE \textbf{Initialize}: $\mathbf{a},\mathbf{w},\{\tilde{\mathbf{r}}_n^{(0)}\}_{n=1}^N,\{\tilde{\mathbf{v}}_n^{(0)}\}_{n=1}^N$.
		
		\STATE  Evaluate the fitness value for each particle via \eqref{eq19}.
		
		\STATE \textbf{for} $t=1\rightarrow T$ \textbf{do} (Outer Loop)
		
		\STATE \hspace{0.25cm} \textbf{for} $n=1\rightarrow N$ \textbf{do}
		
		\STATE \hspace{0.5cm} Update $\tilde{\mathbf{v}}_n^{(t)}$ and $\tilde{\mathbf{r}}_n^{(t)}$ via \eqref{eq20}
		and \eqref{eq21}, respectively.
		
		\STATE \hspace{0.5cm} \textbf{repeat} (Inner Loop)
		
		\STATE \hspace{0.75cm} Update the receive combining vector $\mathbf{w}$ via \eqref{eq13}.
		
		\STATE \hspace{0.75cm} Update the transmit coefficient vector $\mathbf{a}$ via \eqref{eq15}.
		
		\STATE \hspace{0.5cm} \textbf{until} The decrement on CMSE is smaller than $\epsilon$.
		
	    \STATE \hspace{0.5cm}  Calculate the fitness value $\mathcal{F}\left(\tilde{\mathbf{r}}_n^{(t)}\right)$ via \eqref{eq19}.
	    
	    \STATE \hspace{0.5cm} Update $\tilde{\mathbf{r}}_{n,pbest}=\arg\min\limits_{\tilde{\mathbf{r}}_n}\{\mathcal{F}\left(\tilde{\mathbf{r}}_n^{(t)}\right)),\mathcal{F}\left(\tilde{\mathbf{r}}_{n,pbest}\right)$\}.
	    
	    \STATE \hspace{0.5cm} Update $\tilde{\mathbf{r}}_{gbest}=\arg\min\limits_{\tilde{\mathbf{r}}_n}\{\mathcal{F}\left(\tilde{\mathbf{r}}_n^{(t)}\right)),\mathcal{F}\left(\tilde{\mathbf{r}}_{gbest}\right)$\}.
		
		\STATE \hspace{0.25cm} \textbf{end for}
		
		\STATE \textbf{end for}
		
		\STATE Obtain the APV $\tilde{\mathbf{r}}=\tilde{\mathbf{r}}_{gbest}$ and its corresponding  $\mathbf{w}$ and $\mathbf{a}$.		

		\STATE \textbf{return} $\mathbf{a},\mathbf{w},\tilde{\mathbf{r}}$.
	\end{algorithmic}
	\label{alg1} 
\end{algorithm}

\section{Numerical results}
In this section, we provide numerical results to evaluate the performance of our proposed scheme. In the simulation, we set $A=3\lambda,~L_k=L=5,1\leq k\leq K,~D=\lambda/2$ and $\sigma^2=-80~\mathrm{dBm}$. In addition, the PSO parameters are set to $c_1=c_2=1.5,~\omega_{\max}=0.9,~\omega_{\min}=0.4,~\tau=20,~N=200$ and $T=200$. The PRV of the $k$-th user, i.e., $\mathbf{g}_k,1\leq k \leq K$, is modeled as a circularly symmetric complex Gaussian random vector, with independent and identically distributed (i.i.d.) components $g_{k,n}\sim\mathcal{CN}(0,d_k^{-\alpha}/L),1\leq n\leq L$, where $\alpha=3.9$ is the path-loss factor, $d_k$ is the distance from user $k$ to the AP, uniformly distributed from 250 to 300 m. 
\begin{figure}[t]
	\centering
	\includegraphics[width=0.45\textwidth]{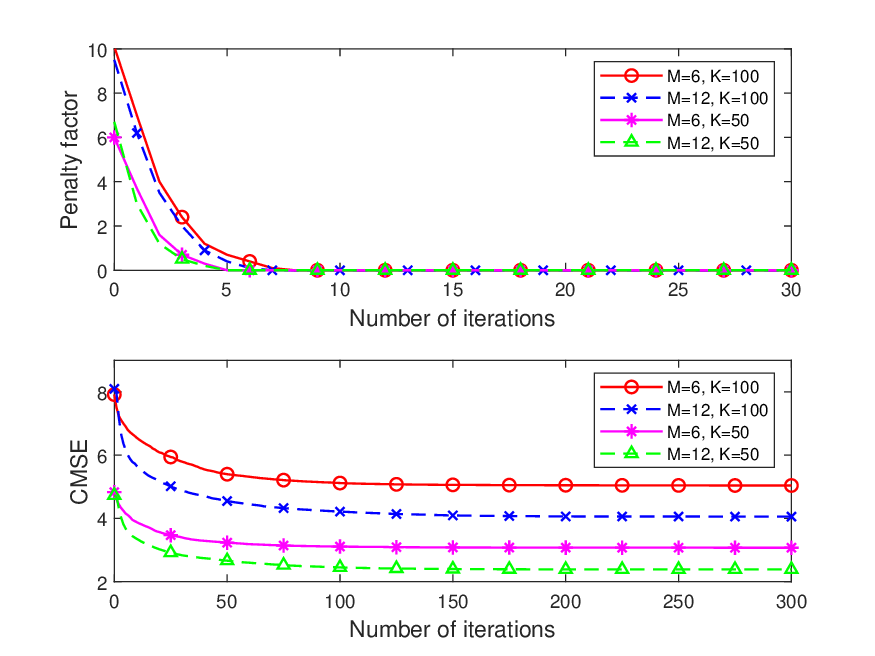} 
	\caption{Convergence performance of our proposed algorithm.}
	\label{convergence_performance}
\end{figure}

For further comparison, we consider three following benchmark schemes: 1) \textbf{FPA}, where the AP is equipped with FPA-based uniform planar array with $M$ antennas; 2) \textbf{AO}, where the three variables of problem \eqref{eq9} are alternately optimized and the optimization of $\tilde{\mathbf{r}}$ are provided in \textbf{Appendix A} by using the SCA; 3) Alternating position selection (\textbf{APS}), where the total moving region of MAs is quantized into discrete locations and each MA's position is alternately selected with the others being fixed \cite{ref4}.
\begin{figure}[t]
	\centering
	\includegraphics[width=0.45\textwidth]{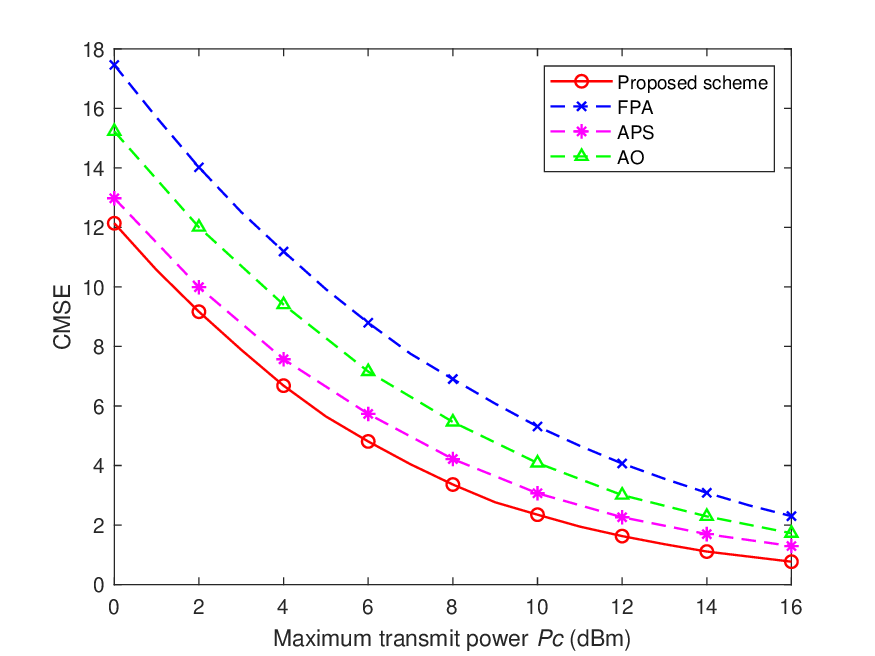} 
	\caption{CMSE for different schemes versus the maximum transmit power.}
	\label{CMSE_vs_Pmax}
\end{figure}

In Fig. \ref{convergence_performance}, we plot the convergence of our proposed algorithm where the maximum transmit power is $P_c=10~\mathrm{dBm}$. The number of MAs at the AP, $M$, and the number of users, $K$, are annotated in the legend. Moreover, to verify the effectiveness of our introduced penalty factor in \eqref{eq19}, we also present the penalty value versus the number of iterations. First, we observe that the penalty value becomes zero after 30 iterations, which guarantees that the minimum inter-MA distance, i.e., constraint \eqref{eq9d}, is satisfied. Then, it is shown that our proposed algorithm converges after about 100 iterations. Besides, we can also find that the achieved CMSE increases with the increase of $K$ or the decrease of $M$.
\begin{figure}[t]
	\centering
	\includegraphics[width=0.45\textwidth]{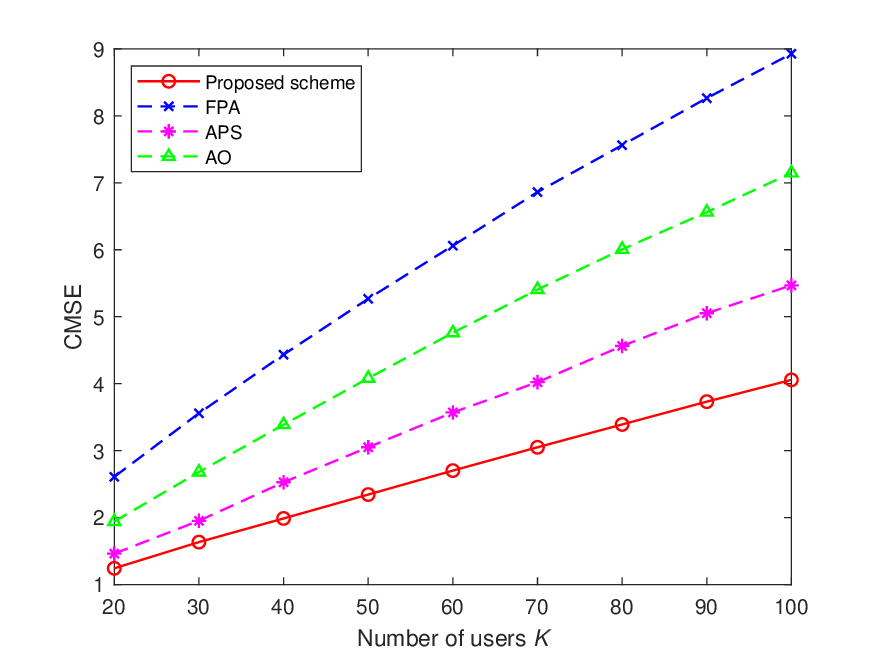} 
	\caption{CMSE for different schemes versus the number of users.}
	\label{CMSE_vs_K}
\end{figure}
\begin{figure}[t]
	\centering
	\includegraphics[width=0.45\textwidth]{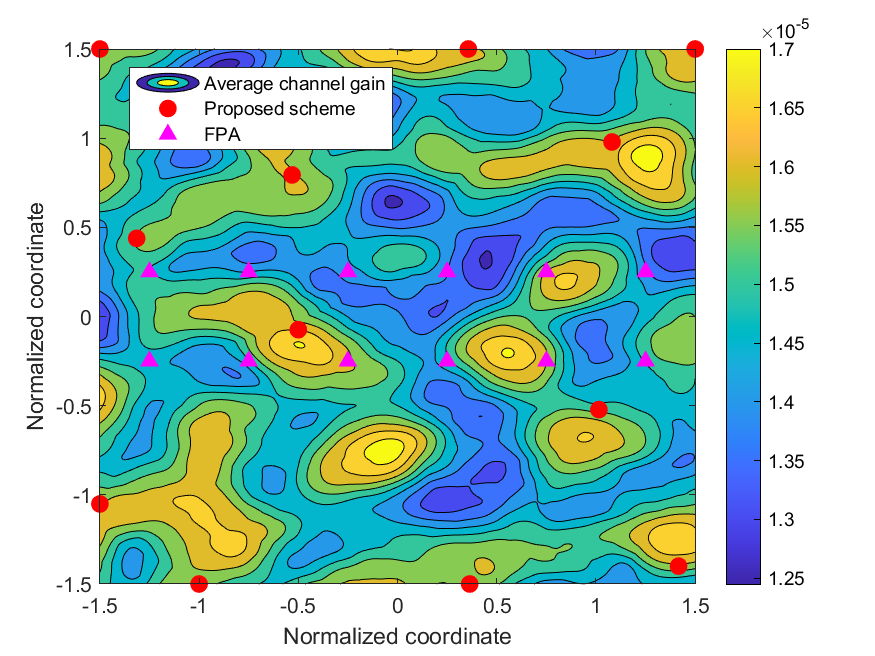} 
	\caption{Average channel gain versus the antenna position.}
	\label{MA_positions}
\end{figure}

In Fig. \ref{CMSE_vs_Pmax}, we show the CMSE versus the maximum transmit power $P_c$, where $M=12$ and $K=50$. As we can see, our proposed scheme significantly outperforms the FPA scheme in reducing the CMSE for any given value of $P_c$. This is because the MA can provide additional DoFs to reduce the CMSE (mainly the first term in \eqref{eq8}) without magnifying the impact of noise (the second term in \eqref{eq8}). In addition, we also observe that our proposed scheme can achieve a lower CMSE than other schemes, such as ``APS'' and ``AO''. However, as $P_c$ increases, the performance gap between our proposed scheme and other benchmark schemes decreases. The reason is that, when the power constraint $P_c$ is large, the users can increase their transmit power to reduce the CMSE caused by the poor channel qualities. As a result, the performance gain provided by MAs will be diminished in such case.
\begin{figure}[t]
	\centering
	\includegraphics[width=0.45\textwidth]{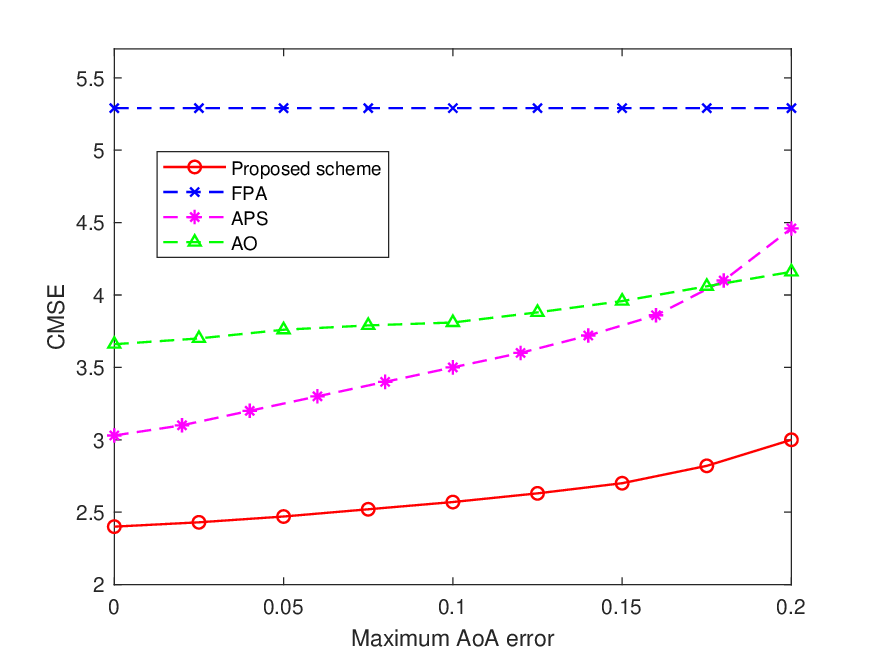} 
	\caption{CMSE for different schemes versus the AoA error.}
	\label{CMSE_vs_error}
\end{figure}

Fig. \ref{CMSE_vs_K} depicts the CMSE versus the user number $K$, where $M=12$ and $P_c=10~\mathrm{dBm}$. As can be observed, the CMSE increases linearly with the number of users, $K$. Meanwhile, our proposed scheme offers substantial improvements than other benchmark schemes and the performance gap increases significantly as $K$ grows. Fig. \ref{MA_positions} shows an example of the optimized MA positions of our proposed scheme for a particular channel realization compared with traditional FPAs, where the average channel gain is defined as $\frac{1}{K}\sum_{k=1}^{K}|h_k|$. This is because, in a multi-user scenario, each user’s channel path has different AoAs and PRVs, such that the MA position optimization should consider the trade-off among the channel gains of all users. From this figure, we can observe that the positions of the MAs are not arranged in a regular manner as the
conventional FPAs. Besides, we observe that in our proposed scheme, the MAs are deployed at
the positions with more favorable average channel gains compared with the FPA system, thereby
enhancing the system performance.

In Fig. \ref{CMSE_vs_error}, we show the impact of the AoA errors on the performance of our proposed scheme, since the FRI errors are generally inevitable in practice. For this purpose, we define the AoA errors as the differences between the estimated AoAs and the actual AoAs, which follow i.i.d. uniformly distributions over $[-\mu/2,\mu/2]$ with $\mu$ representing the maximum AoA error. The system parameters are the same as in Fig. \ref{CMSE_vs_Pmax}. It is shown that the CMSE increases with the maximum AoA error because of the deviation between the estimated and actual channels. However, our proposed scheme can still achieve a significant performance gain compared with other benchmark schemes when $\mu \leq 0.2$.

\section{Conclusion}
In this letter, we considered an MA-aided AirComp network and focused on minimizing the CMSE by jointly optimizing the APV, the receive combining vector, and the transmit coefficient vector. Unlike most related works that adopt the AO or the SCA for handling the resulted highly non-convex problem, we developed an efficient two-loop iterative algorithm based on the PSO. Simulation results shows significant performance gain of our proposed system compared with conventional FPA systems. Future works could be carried out with more practical considerations, such as statistical FRI knowledge, discrete MA locations, and hardware-constrained antenna movement.
{\appendices
	\section{successive convex approximation}
	Notice from \eqref{eq9} that the optimization of $\mathbf{r}_m$ with given $\mathbf{w},\mathbf{a}$ and $\{\mathbf{r}_i,i\neq m\}_{i=1}^{M}$ is expressed as
		\begin{align}
			\label{eq24a}
			\min_{\mathbf{r}_m} ~ \sum_{k=1}^{K}|a_k\mathbf{w}^{\mathrm{H}}\mathbf{h}_k-1|^2,~
			\mathrm{s.t.} ~ \text{\eqref{eq9c},~\eqref{eq9d}}.
		\end{align}
	Then, we expand the above objective function as follows: $\sum_{k=1}^{K}|a_k|^2\mathbf{h}_k^{\mathrm{H}}\mathbf{w}\mathbf{w}^{\mathrm{H}}\mathbf{h}_k-2\mathrm{Re}\left\{a_k^*\mathbf{h}_k^{\mathrm{H}}\mathbf{w}\right\}+1$. Let $w_i$ and $h_{k,i}$ denote the $i$-th entry of $\mathbf{w}$ and $\mathbf{h}_k$, respectively. From \eqref{eq6} and \eqref{eq7}, we know that only the $m$-th entry of $\mathbf{h}_k$ is related to $\mathbf{r}_m$. Then, we can rewrite $\mathbf{h}_k^{\mathrm{H}}\mathbf{w}$ as $\mathbf{h}_k^{\mathrm{H}}\mathbf{w}=w_m\mathbf{g}_k^{\mathrm{H}}\mathbf{f}_k(\mathbf{r}_m)+\beta_k$, where $\beta_k=\sum_{i\neq m}w_ih_{k,i}^*$. Thus, minimizing \eqref{eq24a} is equivalent to minimizing 
	\begin{equation}
		\scalebox{0.9}{$G(\mathbf{r}_m)=\sum_{k=1}^{K}\underbrace{\mathbf{f}_k(\mathbf{r}_m)^{\mathrm{H}}\mathbf{B}_k\mathbf{f}_k(\mathbf{r}_m)}_{G_{k,1}(\mathbf{r}_m)}+2\underbrace{\mathrm{Re}\left\{\mathbf{q}_k^{\mathrm{H}}\mathbf{f}_k(\mathbf{r}_m)\right\}}_{G_{k,2}(\mathbf{r}_m)},$}\label{eq25}
	\end{equation}
	where $\mathbf{B}_k=|a_k|^2|w_m|^2\mathbf{g}_k\mathbf{g}_k^{\mathrm{H}}$ and $\mathbf{q}_k=(|a_k|^2\beta_k-a_k)w_m^*\mathbf{g}_k$. Motivated by \cite{ref4},\cite{ref2}, we aim to find a quadratic surrogate upper bound of \eqref{eq25}. By denoting the $(i,j)$-th entry of $\mathbf{B}_k$ as $|b_{k}^{ij}|e^{\mathrm{j}\angle b_{k}^{ij}}$, $G_{k,1}(\mathbf{r}_m)$ can be further expressed as
	\begin{align}
		\scalebox{0.9}{$
		G_{k,1}(\mathbf{r}_m)$}&\scalebox{0.9}{$=\sum_{i=1}^{L_k}\sum_{j=1}^{L_k}\mathrm{Re}\left\{|b_k^{ij}|e^{\mathrm{j}[\frac{2\pi}{\lambda}(-\rho_k^i(\mathbf{r}_m)+\rho_k^j(\mathbf{r}_m))+\angle b_k^{ij}]}\right\}$}\notag\\
		&\scalebox{0.9}{$=\sum_{i=1}^{L_k}\sum_{j=1}^{L_k}|b_k^{ij}|\cos(\Gamma_k^{ij}(\mathbf{r}_m)),$}
	\end{align}
	where $\Gamma_k^{ij}(\mathbf{r}_m)=\frac{2\pi}{\lambda}(\rho_k^i(\mathbf{r}_m)-\rho_k^j(\mathbf{r}_m))-\angle b_k^{ij}$. Subsequently, we can derive the gradient vector of $G_{k,1}(\mathbf{r}_m)$ as follows:
	\begin{align*}
		\scalebox{0.9}{$\nabla G_{k,1}(\mathbf{r}_m)=\begin{bmatrix}
			\begin{aligned}
				&-\frac{2\pi}{\lambda}\sum_{i=1}^{L_k}\sum_{j=1}^{L_k}|b_k^{ij}|\Upsilon_k(i,j)\sin(\Gamma_k^{ij}(\mathbf{r}_m))\\
				&-\frac{2\pi}{\lambda}\sum_{i=1}^{L_k}\sum_{j=1}^{L_k}|b_k^{ij}|\Phi_k(i,j)\sin(\Gamma_k^{ij}(\mathbf{r}_m))
			\end{aligned}
		\end{bmatrix},$}
	\end{align*}
	where $\Upsilon_k(i,j)=\sin\theta_k^i\cos\phi_k^i-\sin\theta_k^j\cos\phi_k^j$ and $\Phi_k(i,j)=\cos\theta_k^i-\cos\theta_k^j$. Besides, by denoting the $p$-th entry of $\mathbf{q}_k$ as $|q_{k}^{p}|e^{\mathrm{j}\angle q_{k}^{p}}$, we can derive the gradient vector of $G_{k,2}(\mathbf{r}_m)$ as $\nabla G_{k,2}(\mathbf{r}_m)$ from Eq. (40) in \cite{ref4}. Accordingly, the gradient vector of $G(\mathbf{r}_m)$ is given by $\nabla G(\mathbf{r}_m)=\sum_{k=1}^{L_k}\nabla G_{k,1}(\mathbf{r}_m)+2\nabla G_{k,2}(\mathbf{r}_m)$. Then, we further construct a positive number $\xi$ satisfying $\xi\mathbf{I}\succeq\nabla^2 G(\mathbf{r}_m)$. Similar to \cite{ref4}, since $\|\nabla^2 G(\mathbf{r}_m)\|_2\mathbf{I}\succeq \nabla^2 G(\mathbf{r}_m)$, after some manipulations, we have $\xi=\frac{16\pi^2}{\lambda^2}\sum_{k=1}^{K}\left(2\sum_{i=1}^{L_k}\sum_{j=1}^{L_k}|b_k^{ij}|+\sum_{i=1}^{L_k}|q_k^i|\right)$. 
	
	After the aforementioned analysis, by applying the first-order Taylor expansion on $\frac{\xi}{2}\mathbf{r}_m^{\mathrm{T}}\mathbf{r}_m-G(\mathbf{r}_m)$ with given local point $\mathbf{r}_m^i$, we can obtain a surrogate upper bound on $G(\mathbf{r}_m)$ in the $i$-th iteration of SCA, given by
	\begin{align}
		G(\mathbf{r}_m)
		\leq\frac{\xi}{2}\mathbf{r}_m^{\mathrm{T}}\mathbf{r}_m+\left(\nabla G(\mathbf{r}_m^i)-\xi\mathbf{r}_m^i\right)^{\mathrm{T}}\mathbf{r}_m+\Omega,
	\end{align}
	where $\Omega=G(\mathbf{r}_m^i)-(\nabla G(\mathbf{r}_m^i)-\frac{\xi}{2}\mathbf{r}_m^i)^{\mathrm{T}}\mathbf{r}_m^i$ is a constant value. Moreover, according to Eq. (29) in \cite{ref4}, constraint \eqref{eq9d} can be relaxed as
	\begin{equation}
		\scalebox{0.96}{$\frac{1}{\|\mathbf{r}_m^i-\mathbf{r}_n\|_2}(\mathbf{r}_m^i-\mathbf{r}_n)^{\mathrm{T}}(\mathbf{r}_m-\mathbf{r}_n)\geq D, 1\leq n \neq m \leq M.$}\label{eq28}
	\end{equation}
	Hereto, in the $i$-th iteration of SCA, the optimization problem for the $m$-th MA is transformed
	into
	\begin{align}
		\label{eq29}
			\min_{\mathbf{r}_m} ~ \frac{\xi}{2}\mathbf{r}_m^{\mathrm{T}}\mathbf{r}_m+\left(\nabla G(\mathbf{r}_m^i)-\xi\mathbf{r}_m^i\right)^{\mathrm{T}}\mathbf{r}_m,~
			\mathrm{s.t.}~\text{\eqref{eq9c},\eqref{eq28}},
	\end{align}
	 which is a convex optimization problem and can be solved by using the CVX toolbox.
}

\bibliography{reference}
\bibliographystyle{IEEEtran}

\vfill
\end{document}